\documentclass[12pt,preprint]{emulateapj}

\shorttitle{The relative abundance of compact and normal massive
  early-types at $0<z<2.5$} \shortauthors{Cassata et al.}
\def \zmed {$<\! z\!>$}
\begin{document}

\title{The Relative Abundance of Compact and Normal Massive Early-Type
  Galaxies and its evolution From Redshift $z\sim2$ To The Present.
  \altaffilmark{1}}

\altaffiltext{1}{Based on data obtained with the \textit{Hubble Space Telescope} operated by AURA, Inc. for NASA under contract NAS5-26555. }

\author{P. Cassata\altaffilmark{2}, 
M. Giavalisco\altaffilmark{2}, 
Yicheng Guo\altaffilmark{2},
A. Renzini\altaffilmark{3},
H. Ferguson\altaffilmark{4},
A. M. Koekemoer\altaffilmark{4},
S. Salimbeni\altaffilmark{2},
C. Scarlata\altaffilmark{5},
N. A. Grogin\altaffilmark{4},
C. J. Conselice\altaffilmark{6}
T. Dahlen\altaffilmark{4},
J. M. Lotz\altaffilmark{4},
M. Dickinson\altaffilmark{7},
and Lihwai Lin\altaffilmark{8}
}

\altaffiltext{2}{Department of Astronomy, University of Massachusetts, Amherst, MA 01003; paolo@astro.umass.edu}
\altaffiltext{3}{Osservatorio Astronomico di Padova (INAF-OAPD), Vicolo dell'Osservatorio 5, I-35122, Padova Italy}
\altaffiltext{4}{Space Telescope Science Institute, 3700 San Martin Boulevard, Baltimore, MD, 21218}
\altaffiltext{5}{School of Physics and Astronomy, University of Minnesota, 116 Church Street S.E., Minneapolis, MN 55455}
\altaffiltext{6}{University of Nottingham, School of Physics and Astronomy, Nottingham NG7 2RD}
\altaffiltext{7}{NOAO-Tucson, 950 North Cherry Avenue, Tucson, AZ 85719}
\altaffiltext{8}{Institute of Astronomy \& Astrophysics, Academia Sinica, Taipei 106, Taiwan}

\begin{abstract}
We report on the evolution of the number density and size of
early-type galaxies from $z\sim2$ to $z\sim0$. We select a sample of
563 massive ($M>10^{10}M_{\odot}$), passively evolving ($SSFR<10^{-2}
Gyr^{-1}$) and morphologically spheroidal galaxies at $0<z<2.5$, using
the panchromatic photometry and spectroscopic redshifts available in
the GOODS fields. We combine ACS and WFC3 HST images to study the
morphology of our galaxies in their optical rest-frame in the whole
$0<z<2.5$ range. We find that throughout the explored redshift range
the passive galaxies selected with our criteria have weak
morphological K-correction, with size being slightly smaller in the
optical than in the UV rest-frame (by $\sim$20\% and $\sim$10\% at
$z>1.2$ and $z<1.2$, respectively). We measure a significant evolution
of the mass-size relation of early-type galaxies, with the fractional
increment that is almost independent on the stellar mass. Early-type
galaxies (ETGs) formed at $z>1$ appear to be preferentially small, and
the evolution of the mass-size relation at $z<1$ is driven by both the
continuous size growth of the compact galaxies and the appearance of
new ETGs with large sizes. We also find that the number density of all
passive early-type galaxies increases rapidly, by a factor of 5, from
$z\sim2$ to $z\sim1$, and then more mildly by another factor of 1.5
from $z\sim1$ to $z\sim0$. We interpret these results as the evidence
that the bulk of the ETGs are formed at $1<z<3$ through a mechanism
that leaves very compact remnants. At $z<1$ the compact ETGs grow
gradually in size, becoming normal size galaxies, and at the same time
new ETGs with normal-large sizes are formed.

\end{abstract}


\section{Introduction}
The epoch between $z\sim3$ and $z\sim1$ appears to be crucial for the
assembly of massive galaxies, as many authors find that the number
density of galaxies with masses around 10$^{11} M_{\odot}$ and larger
evolves very rapidly over this period of time (Fontana~et~al.~2006;
Marchesini~et~al.~2009; Ilbert~et~al.~2010; Conselice~et~al.~2011). By
redshift $z\sim1$, however, the number density of early-types is
comparable to the local one, with an increase of a factor of 1.5
allowed at the most (Cimatti~et~al.~2006; Franceschini~et~al.~2006;
Borch~et~al.~2006). There is also evidence that a substantial fraction
of the massive elliptical galaxies at $z>1.5$ are smaller than their
local counterparts of the same mass (Daddi~et~al.~2005;
Trujillo~et~al.~2007; Toft~et~al.~2007; Zirm~et~al.~2007;
Van~Dokkum~et~al.~2008; Buitrago~et~al.~2008; Cimatti~et~al.~2008),
whereas some passively evolving galaxies may be already present at
redshifts as high as $\sim 3$ (Van~Dokkum~et~al.~2010; Guo
et~al.~2011a in prep.).

We still lack a comprehensive understanding of the formation of
passively evolving galaxies and of their evolution in size and spatial
abundance from $z\sim 3$ to the present, including the physical
mechanisms responsible for the quenching of star formation.
Classically, hierarchical models of galaxy formation predict that the
evolution of massive galaxies at early epochs is mostly driven by
major mergers (e.g. Shankar~et~al.~2010, 2011) and that some form of
feedback, from star formation itself or from AGN activity, quenches
star formation (e.g. Somerville~et~al~2008). But highly turbulent
disks with SFR $\sim100 M_{\odot}yr^{-1}$ have been claimed to also be
efficient into slowing accreting gas in recent works
(F\"{o}rster-Schreiber~et~al.~2006; F\"{o}rster-Schreiber~et~al.~2009;
Genzel~et~al.~2008). Others have suggested that at later epochs
($z<1$), the evolution is dominated by less violent processes, like
minor mergers and/or smooth accretion (Van~Dokkum~et~al.~2010), that
can possibly explain the size evolution as well
(Hopkins~et~al.~2009). Recently, Peng et~al.  (2010) have shown that
two independent processes are responsible for the quenching, one
related to mass (or almost equivalently to the star formation rate),
and another to the environment via the local overdensity, with the
former dominating especially at high masses and early cosmic times,
whereas the latter dominates at low masses and late times.

Observationally, there is still ongoing debate on the quantitative and
qualitative features of the evolution of early-type galaxies. For
example, the detection of compact size, and thus large stellar
density, is mostly based on morphological studies done either using
{\it HST}/ACS $z$-band images, which at $1<z<2$ sample the rest-frame
UV and thus might not reflect the morphology of the bulk of the
stellar mass (Daddi~et~al.~2005; Trujillo~et~al.~2007;
Cimatti~et~al.~2008); or on {\it HST}/NICMOS or ground--based near-IR
images, which have substantially worse resolution and/or image quality
than ACS, and thus are potentially affected by systematic errors
(Toft~et~al.~2007; Zirm~et~al.~2007; Van~Dokkum~et~al.~2008;
Buitrago~et~al.~2008). Even when taking advantage of modern,
high-quality near--IR imagers such as {\it HST}/WFC3, only quite small
samples have been studied so far (Cassata~et~al.~2010;
Szomoru~et~al.~2010).

One essential step to map the evolution of passively evolving galaxies
is the measure of the distribution function of their size at various
cosmic epochs. While such distribution function has been studied in
the local universe, at $z\sim 2$ the measures remain uncertain. There
is evidence that the size of high-redshift early-type galaxies (ETG)
covers a range that overlaps with those in the present-day universe
(e.g. Saracco et~al. 2010), whereas selection effects might contribute
to underestimate the contribution by more extended galaxies in the
high-z samples (Mancini et~al. 2010). In any case, the mass-size
relation of passively evolving galaxies and its evolution with
redshift still remains poorly constrained.  Even at $z\sim0$, the
distribution function of the effective radius of early-type galaxies
is not well characterized at the small end, with some groups
suggesting that a compact population of early-types could be hidden
among unresolved red objects (Shih~\&~Stockton~(2011). Moreover,
Valentinuzzi~et~al.~(2010a; 2010b) reported evidence of a rich
population of compact passive galaxies at $z\sim0.04$ and $z\sim0.7$,
and Saracco~et~al.~(2010) pointed out that the number density of
$z\sim2$ compact galaxies is comparable with the local density of
similarly small early-types.

Recently, various groups have analyzed the morphology of massive
early-type galaxies at $z\sim 2$ using WFC3 images in the HUDF and/or
ERS field (both located within the boundaries of GOODS-South) and
found that the size of passive galaxies at $z\sim2$ is almost
independent of the wavelength, confirming previous results on the
compactness of such galaxies (e.g., Cassata et~al. 2010;
Ryan~et~al. 2011; Szomoru et~al. 2010). If indeed the rest-frame UV
morphology of these galaxies is similar to that at optical wavelength
this allows us to use the wider ACS database for statistically
significant studies of the distribution function of their effective
radius, its dependence with the stellar mass, and its evolution with
redshift.

The aim of this paper is to compare the size distribution {\it in the
  same rest-frame optical} for homogeneous samples of early-type
galaxies at $0<z<2.5$, selected in a consistent way, and to constrain
the number evolution of normal and compact early-types over the same
epoch. In particular, we exploit the GOODS multi-band dataset to build
a robust sample of passive early-type galaxies at $1.2<z<2.5$, and a
similar sample of galaxies at $0<z<1.2$. We complement those two with
a sample of local early-type galaxies from the SDSS survey, selected
using the same criteria. Using the SDSS sample, we establish the local
mass-size relation, and we identify local compact ETGs. Classically,
authors in literature consider compact any galaxies $1-\sigma$ below
the local SDSS relation (Valentinuzzi~et~al.~2010a,
Valentinuzzi~et~al.~2010b, Saracco~et~al.~2010). This definition of
compactness implies that 16\% of the local galaxies are {\it
  compact}. In this paper, we use this definition, but we identify as
well a class of {\it ultra-compact} early-type (at all redshifts) with
size 0.4 dex below the local average relation. Virtually, there are no
such dense galaxies in the local Universe. With the aim of
constraining the evolutionary path that early-type galaxies follow
from $z\sim2$ to the present, we compute the mass-density in {\it
  normal}, {\it compact} and {\it ultra-compact} early-types at
$0<z<2.5$.

The issue that we try to clarify in this study is to which extent the
evolution with cosmic time of the average mass-size relation is driven
by the size growth of individual galaxies or by the progressive
appearance of ETGs with larger size, as suggested by Valentinuzzi et
al. (2010b).

Throughout the paper, we use a standard $\Lambda$CDM cosmology, with
$\Omega_{\Lambda}$=0.7, $\Omega_{M}$=0.3 and $H_0=70$~Km~s$^{-1}$
~Mpc$^{-1}$, and we assume a Salpeter IMF (Salpeter~1955).

\section{Data and sample selection}
\begin{figure}
\begin{center}
\includegraphics[width=\columnwidth]{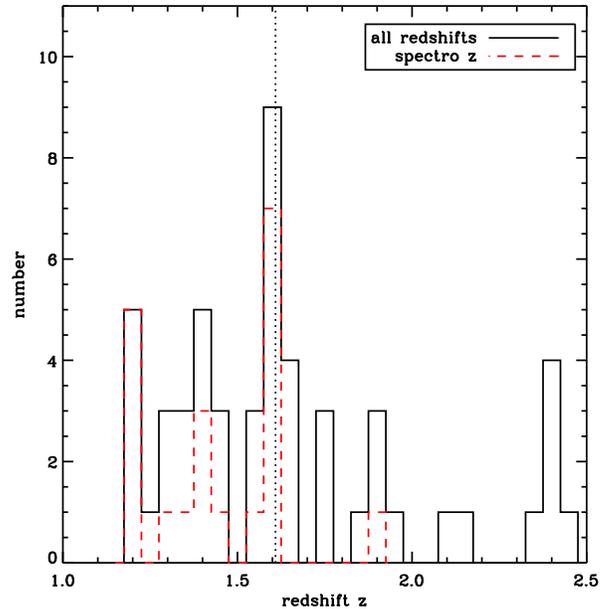}
\caption{Redshift distribution in the highest redshift bin
  $1.2<z<2.5$. The continuous black line and the dashed one indicates
  all objects and objects with spectroscopic redshift,
  respectively. The vertical dotted line at $z\sim1.6$ indicates the
  median of the redshift distribution (that does not change
  considering spectroscopic objects only).}\label{fig1}
\end{center}
\end{figure}

\begin{figure}
\begin{center}
\includegraphics[width=\columnwidth]{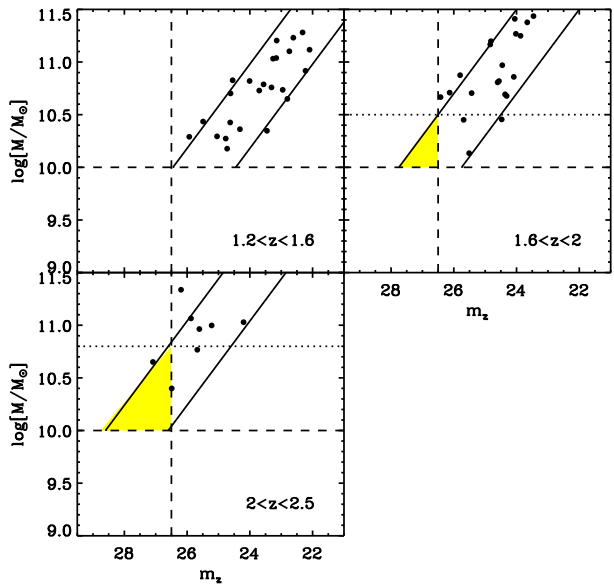}
\caption{AB magnitude in the ACS/$z$-band band versus stellar mass,
for galaxies in our sample at $z>1.2$, divided in three smaller
redshift bins ($1.2<z<1.6$, $1.6<z<2$ and $2<z<2.5$). The vertical
dashed line at $m_z=26.5$ indicates the magnitude at which the GOODS
z-band images are 80\% complete, the dashed orizontal line at
$M/M_{\odot}=10^{10}$ shows our mass limit, and the dotted orizontal
lines at $M/M_{\odot}=10^{10.5}$ and $M/M_{\odot}=10^{10.8}$ show the
limit above which the $1.6<z<2$ and $2<z<2.5$ are complete,
respectively. The diagonal lines indicate, for each redshift bin, the
mass vs $m_z$ relation for the reddest and bluest models available in
our template grid. The yellow triangles at $z>1.6$ show the regions of
the plane in which our selection is incomplete.  }\label{fig2}
\end{center}
\end{figure}

The GOODS-North and GOODS-South, centered on the Hubble Deep Field
North (HDFN) and the Chandra Deep Field South (CDFS), respectively,
provide an unique resource of data to study the distant universe. The
two fields cover a total area of about 300 arcmin$^2$, and have now
been imaged with all the largest available telescopes (Hubble,
Spitzer, VLT, Chandra, XMM, Herschel, CFHT).  The GOODS HST Treasury
Program (Giavalisco~et~al.~2004) provides ultradeep high-resolution
images in B-, V-, i- and z-bands. Deep ground-based imaging in the
U-band is provided by VIMOS/VLT for the CDFS (Nonino~et~al.~2009), and
at the Kitt Peak National Observatory for the HDFN. Moreover,
VLT/ISAAC imaged the CDFS in J-, H- and K-bands, while CFHT/WIRCAM
imaged the HDFN in J- and K-bands (Lin et al.  in prep; also see Wang
et al. 2010).  Ultradeep Spitzer/IRAC imaging is also available in the
3.6, 4.5, 5.6 and 8.0 $\mu$m NIR channels.

We built a multiwavelength catalog (GUTFIT, Grand Unified TFIT
catalog) for each of the GOODS-N and GOODS-S using the TFIT procedures
by Laidler~et~al.~(2007). This procedure allows to PSF match images
having different resolutions, and uses the ACS/$z-$band (version 2.0)
as detection image. The positions and profiles of galaxies in the
ACS/$z-$band high resolution image are used as a prior to reconstruct
the flux in the lower resolution images. This is useful for objects
that are deblended in the high resolution images but are close enough
in the sky to overlap in the low resolution ones. In this cases, TFIT
is able to reconstruct the flux of each object, and is accurate down
to the limiting sensitivities of images (Laidler~et~al.~2007)


We included all the WFC3 data available in the GOODS-South field to
July 2010, namely the HUDF observations (GO 11563, PI: Illingworth)
and the Early Release Science Program 2 (ERS2: GO 11359. PI:
McConnell; see Windhorst~et~al.~2011 for further details). The former
observed one WFC3 field ($\sim4$~arcmin$^2$) centered in the HUDF in
the F105W ($Y$), F125W ($J$) and F160W ($H$) filters, imaged
respectively for 16, 16 and 28 orbits, reaching $1-\sigma$
fluctuations of 27.2, 26.6 and 26.3 $AB/arcsec^2$ in the three bands,
respectively. The latter covers 40~arcmin$^2$ on the north part of the
GOODS South field, imaged with the same filters as the HUDF/WFC3, with
integration times of 2, 2 and 3 orbits respectively for the F105W
($Y$), F125W ($J$) and F160W ($H$) filters, producing $1-\sigma$
fluctuations of 25, 24.4 and 24.1 $AB/arcsec^2$ in the three bands,
respectively. We used a version of these datasets that had been
processed as described in more detail in Koekemoer~et~al.~(2011),
which were combined using MultiDrizzle (Koekemoer~et~al.~2002) to a
0.06'' pixel scale, aligned to the existing GOODS and HUDF astrometric
grid, and obtaining a PSF of $\sim$0.16'' in our resulting WFC3
images.

The sample was complemented with the 6000 spectroscopic redshifts
available to date (3000 in the South and 3000 in the North), among
which 2500 are at $z>1$ (Cimatti~et~al.~2008; Vanzella~et~al.~2008;
Popesso~et~al.~2009).


\begin{figure*}
\begin{center}
\includegraphics[width=\textwidth]{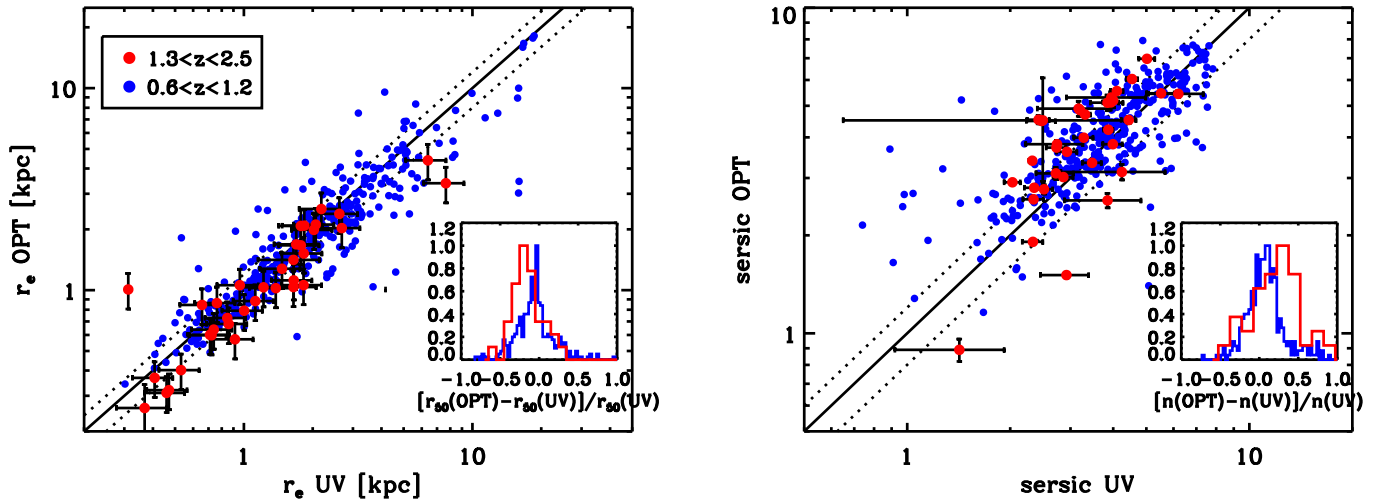}
\caption{Comparison between the GALFIT best fit values in the optical
  and UV rest-frame for our sample of passive galaxies at $z\sim2$
  (red) and the sample of passive galaxies at $0.6<z<1.2$ (blue). The
  left and right panels show the half light radii and the S\'ersic
  indices, respectively. The dotted lines show the $\pm$20\%
  lines. The two insets show the distribution of the fractional
  differences between sizes and S\'ersic indices in the optical and UV
  rest-frames.}\label{fig3}
\end{center}
\end{figure*}

We fitted the Spectral Energy Distribution of the galaxies in the two
samples to measure accurate photo-z, using the PEGASE 2.0 templates
(Fioc~\&~Rocca-Volmerange 1997). By comparing photo-z and spectro-z, we
estimate that the scatter of our photo-z is of the order of
$\sigma[\Delta z/(1+z)]\sim0.04$.

Once a redshift has been assigned to each galaxy (either spectroscopic
or photometric), we fitted the SEDs from the UV to 8$\mu$m to a set of
models by Charlot~\&~Bruzual 2009 (CB09, in preparation), in order to
get accurate measurements of the stellar mass, E(B-V), age and Star
Formation Rate of sources. In particular, we use a Salpeter IMF
(Salpeter~1955) with lower and upper mass of 0.1 and 100 $M_{\odot}$,
we apply the Calzetti law (Calzetti~et~al.~2000) to describe the dust
extinction and we use exponentially declining Star Formation
Histories. Maraston~et~al.~(2010) showed that these models tend to
overestimate the SFRs and underestimate stellar masses for a sample of
star-forming galaxies at $z\sim$2, and that exponentially increasing
SFH give instead less biased results. We will explore if such biases
are present for early-type galaxies as well in a forthcoming
paper. The stellar mass that we derive does not include the the mass
loss from stars, but includes the stellar mass of remnants (white
dwarfs, neutron stars, etc..). The Specific Star Formation Rates
(SSFR), i. e. the star formation rate for unit mass, are derived by
dividing the star formation rates by stellar masses. The stellar
masses (and all the derived quantities) based on different IMFs that
we cite and use as a comparison throughout the paper were homogenized
to match our assumptions, using the offsets measured by
Salimbeni~et~al.~(2009): $\log$(M$_{CB09}$)=$\log$(M$_{BC03})-0.2$ and
$\log$(M$_{Salp}$)=$\log$(M$_{Chab})+0.25$, where BC03 stands for
Bruzual~\&~Charlot(2003). This process of homogenization is essential
to ensure meaningful comparison among different samples at different
redshifts.

We then extracted our sample of passive galaxies, selecting only
objects with stellar mass $M >10^{10}\,M_{\odot}$ and SSFR$<10^{-2}
Gyr^{-1}$ at $0<z<2.5$. These criteria ensure that only the most
massive and less star-forming objects at those redshifts are selected.
This selection has resulted in a sample of about 900 candidate ETGs at
redshifts between $\sim 0.1$ and $\sim 2.5$. As we want to have an
optical high-resolution image for each candidate, we kept only those
$z>1.2$ galaxies that lie within either the ERS or HUDF WFC3 areas.





We visually classified both the $z<1.2$ and $z>1.2$ sample, in the
$z-$ and $H-$band respectively, and we kept only galaxies with a
spheroidal morphology, i.e. galaxies with no signs of asymmetry and
centrally concentrated. About 60\% of all passive galaxies were
classified as morphologically spheroidals, and were kept in the
sample. We also checked the Spitzer/MIPS images, and we found a
negligible fraction of objects detected at 24$\mu$m ($\sim$10 objects
at $z<1.2$ and 2 at $z>1.2$) that we removed from the sample. At the
end, we end up with 563 passive galaxies, 52 of which are at $z>1.2$.
We stress once again that the 511 early-types at $z<1.2$ are extracted
from the 300 arcmin$^2$ of GOODS-S\&N, while the 52 at $1.2<z<2.5$ are
extracted from the ERS and HUDF WFC3 fields (for a total area of 49
arcmin$^2$).  Figure~\ref{fig1} shows the redshift distribution in the
highest redshift bin, $1.2<z<2.5$, peaked around $z\sim1.6$. 20
galaxies among the 52 in this redshift range have spectroscopic
redshifts.

In Figure~\ref{fig2} we show the stellar mass as a function of
the ACS/$z$-band magnitude for galaxies at $z>1.2$, with the aim of
assessing the incompleteness of our selection at those redshifts. For
three redshift bins ($1.2<z<1.6$, $1.6<z<2$ and $2<z<2.5$) we compare
the stellar mass vs $m_z$ of objects in the sample with the models
used to fit their SEDs. By definition, we can detect in our sample
only galaxies falling between the two diagonal lines in
Fig~\ref{fig2}, that represent the bluest and reddest models. We used
a Montecarlo approach, simulating galaxies with deVaucouleurs profiles
and different sizes, to establish the detection limit of the $z-$band
imaging version 2.0 that is used by GUTFIT as detection image. The
objects are then placed on the real $z-$band images, and the same
SExtractor procedure used to extract real objects is run on the
simulated images. Ignoring for now the dependence of the completeness
on the size of the galaxies (that will be discussed in section~4), we
find that our selection achieves 80\% completeness at $m_z$=26.5, for
compact objects. At $1.2<z<1.6$ the mass vs $m_z$ relation for the
reddest model and the 80\% completeness magnitude intersect around
$M/M_{\odot}=10^{10}$: this implies that up to $z\sim1.6$ we detect
all passive galaxies with any color down to $M/M_{\odot}=10^{10}$. At
higher redshift the model and the 80\% completeness line intersect at
higher masses, implying that our selection misses part of the galaxies
with $M/M_{\odot}>10^{10}$. In particular, at $1.6<z<2$ and at
$2<z<2.5$ our sample is complete only down to $M/M_{\odot}>10^{10.5}$
and $M/M_{\odot}>10^{10.8}$, respectively.

For both the samples we used the GALFIT package (Peng~et~al.~2002)
to fit the light profile in the rest-frame UV and optical to the
S\'ersic model

\begin{equation}
I(r)=I_eexp\left\{-b_n\left[\left(\frac{r}{r_e}\right)^{1/s}-1\right]\right\},
\end{equation}

where $I(r)$ is the surface brightness measured at distance $r$, $I_e$
is the surface brightness at the effective radius $r_e$ and $b_n$ is a
parameter related to the S\'ersic index $s$. For $s$=1 and $s$=4 the
S\'ersic profile reduces respectively to an exponential and
deVaucouleurs profile. Bulge dominated objects typically have high $s$
values (e.g. $s>2$) and disk dominated objects have $s$ around
unity. Ravindranath~et~al.~(2006), Cimatti~et~al.~(2008) and
Trujillo~et~al.~(2007) showed that GALFIT yields unbiased estimates of
the S\'ersic index and effective radius for galaxies with S/N$>10$ and
$r_e>0.03'',$ independently of the redshift of the source, thus
demonstrating that the surface brightness dimming is not an issue for
this kind of studies.

The PSF was obtained in each passband needed by averaging
well-exposed, unsaturated stars. We run GALFIT experimenting on
various sizes of the fitting region around each galaxy, and with the
sky either set to a pre-measured value or left as a free parameter. We
verified that the sizes and S\'ersic indices do not vary by more than
10\% in the various cases. The values that we show throughout the
paper were obtained with a free sky and $6\times6$ arcsec$^2$ fitting
regions. Any closeby object detected by SExtractor within each fitting
region was automatically masked out during the fitting procedure.

Finally, we selected a sample of local galaxies from the Sloan Digital
Sky Survey (SDSS), for which masses, star formation rates and
morphological parameters are available in literature. In particular,
we combined the MPA SDSS DR7 catalog, that contains stellar masses and
star formation rates (Kauffmann~et~al.~2003; Brinchmann~et~al.~2004)
with the DR7 NYU Value-Added Galaxy Catalog, that contains the GALFIT
S\'ersic best fit to the $u$, $g$, $r$, $i$ and $z$ SDSS images
(Blanton~et~al.~2005). We defined the local sample of massive and
passive early-type galaxies by applying the same criteria used for the
high-$z$ ones: stellar mass $M_{\odot}>10^{10}$ and specific star
formation rate SSFR$<10^{-2} Gyr^{-1}$. Instead of visually inspecting
the whole sample, we eliminated the disk dominated objects removing
all the objects (5\% of the total) with S\'ersic index $s<2$ in the
$r-$band. We verified with a random sample of 200 SDSS galaxies that
the contamination of disk dominated objects among objects with $s>2$
is below 5\%.


\section{Morphological properties as a function of the rest-frame band}\label{sect:morphok}
In Figure~\ref{fig3} we compare the S\'ersic indices and the physical
sizes in the UV and optical rest-frames, for the high and low redshift
samples. Since the errors on the sizes and S\'ersic indices given by
GALFIT are the formal errors derived by the fitting procedure and do
not take into account any systematic uncertainty, in this Figure and
in the following we set the error bars to a minimum value of 20\%. For
the low redshift sample we restricted the analysis at $z>0.6$, and we
used the ACS $v-$band and $z-$band, matching the rest-frame UV and
optical, respectively. We did not include galaxies at lower redshift,
as the UV rest-frame at $z<0.6$ would be matched by the ACS/B-band,
that provides a worse S/N than the other ACS bands. As we already said
in the previous section, the $z\sim2$ sample contains the 52 galaxies
for which the WFC3 imaging is available: at that redshift, the $z-$
and the $H$-band correspond respectively to the rest-frame UV at
$\sim$3000\AA~and optical at $\sim$5700\AA. First of all, 95\% of the
passive galaxies at $z<1.2$ and $z\sim2$ have a S\'ersic index $s>2$,
both at UV and optical wavelengths. This is not surprising, as the
sample has been restricted to contain only visually classified
spheroidals. Secondly, the sizes measured in the UV and in the optical
correlate almost perfectly with each other, with a scatter smaller
than $\pm$20\%. The S\'ersic indices in the two rest-frame bands are
correlate as well, but show a larger scatter ($\sim40$\%). However, we
do see systematic offsets for both sizes and S\'ersic indices in the
two rest-frames, that become more evident when checking the fractional
differences between optical and UV, shown in the insets of
Figure~\ref{fig3}. We see that on average galaxies at $1.2<z<2.5$ have
sizes 20\% smaller and S\'ersic indices 25\% larger in the optical
than in the UV rest-frame. Similarly, galaxies at $0.6<z<1.2$ galaxies
have sizes 10\% smaller and S\'ersic indices 10\% larger in the
optical than in the UV rest-frame.

This reinforces our previous result presented in
Cassata~et~al.~(2010), where we showed a similar trend for the subset
of galaxies at $z\sim2$ lying in the HUDF, and it is in good agreement
with other studies of early-type galaxies at $z\sim2$
(McCarthy~et~al.~2007; Trujillo~et~al.~2007; Ryan~et~al.~2011), as
well as at lower redshift (Papovich~et~al.~2003; Cassata~et~al.~2005).

Interestingly, these biases basically imply that early-type galaxies
at $z>1.2$ show a negative color gradient, with the center being
redder than the outskirts. Early-types at lower redshift show a
similar, but shallower, color gradient. This is in very good agreement
with the results by Guo~et~al.~(2011b), who found negative color
gradients for 6 passive galaxies in HUDF, steeper than those observed
in local early-type galaxies.

Based on these evidences, we can conclude that in the whole $0<z<2.5$
interval the morphological K-correction is weak for passive
spheroidals that have already ended their star formation activity.


\section{The mass-size relation for early-type galaxies at $0<z<2.5$}
\begin{figure}
\begin{center}
\includegraphics[width=\columnwidth]{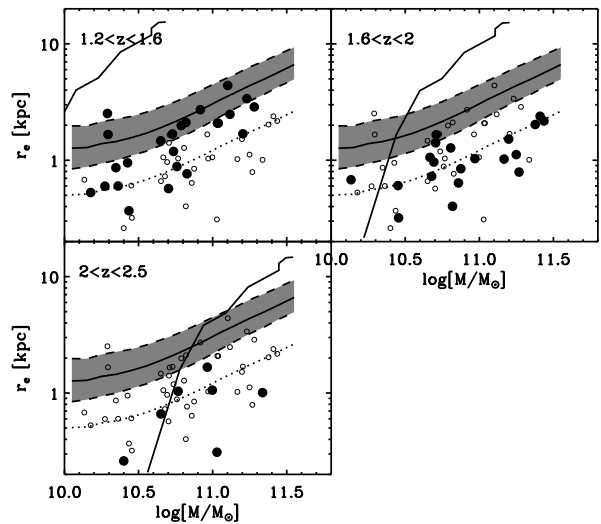}
\caption{The mass size relation in the optical rest-frame for the
  galaxies at $z>1.2$ extracted from the HUDF+HUDF WFC3 fields, in
  three redshift bins: $1.2<z<1.6$, $1.6<z<2$ and $2<z<2.5$. The black
  filled circles indicate objects in that redshift bin, while empty
  circles represent galaxies in the other two redshift bins, for
  comparison. In all panels, the gray filled region indicates the
  locus occupied by SDSS passive galaxies at $0<z<0.1$: the continuous
  line shows the median of the distribution, while the dashed black
  lines contain 68\% of the objects. The black dotted line corresponds
  to the locus of galaxies 0.4 dex smaller than local counterparts:
  galaxies below the line are considered ultra-compact according to
  our definition. The black continuous curves encompass the region of
  the plane where our selection detects at least 80\% of the
  galaxies. Galaxies left of the line have surface brightnesses too
  low to be detected.}\label{fig4}
\end{center}
\end{figure}

\begin{figure}
\begin{center}
\includegraphics[width=\columnwidth]{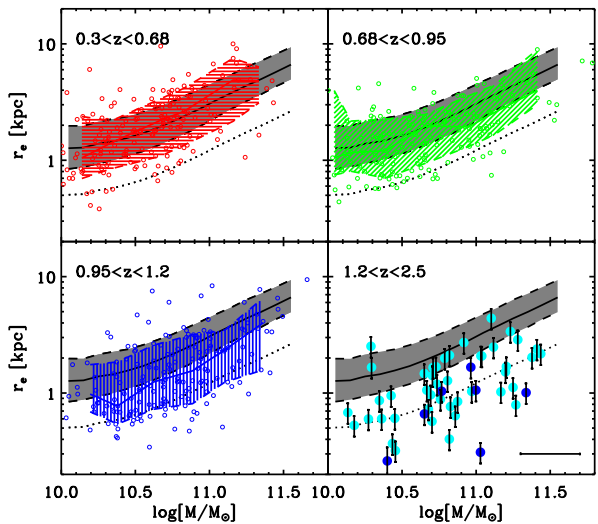}
\caption{The mass size relation in the optical rest-frame for in four
  redshift bins: $0.3<z<0.68$, $0.68<z<0.95$, $0.95<z<1.2$ and
  $1.2<z<2.5$. For the first three bins we used the $z-$ACS band,
  while for the last we used the $F160W$-WFC3 band. The first three
  bins are designed to contain roughly the same number of galaxies. In
  all bins, the gray filled region indicates the locus occupied by
  SDSS passive galaxies at $0<z<0.1$: the continuous line shows the
  median of the distribution, while the dashed black lines contain
  68\% of the objects. The black dotted line corresponds to the locus
  of galaxies 0.4 dex smaller than local counterparts: galaxies below
  the line are considered ultra-compact according to our
  definition. In the first three panels the colored regions indicate
  the distribution of the passive galaxies in each redshift bin: the
  continuous colored line indicates the median of the distribution,
  and the dashed lines contain 68\% of the galaxies. The empty color
  circles indicate individual galaxies. For the highest redshift bin,
  we show each galaxy with its error bar on the physical size: cyan
  and blue points indicate galaxies at $1.3<z<2$ and $2<z<2.5$,
  respectively. The error bar at $\log[M/M_{\odot}]=11.5$ and
  $r_e=0.2 kpc$ shows the typical error on the mass
  estimates.}\label{fig5}
\end{center}
\end{figure}

\begin{figure}
\begin{center}
\includegraphics[width=\columnwidth]{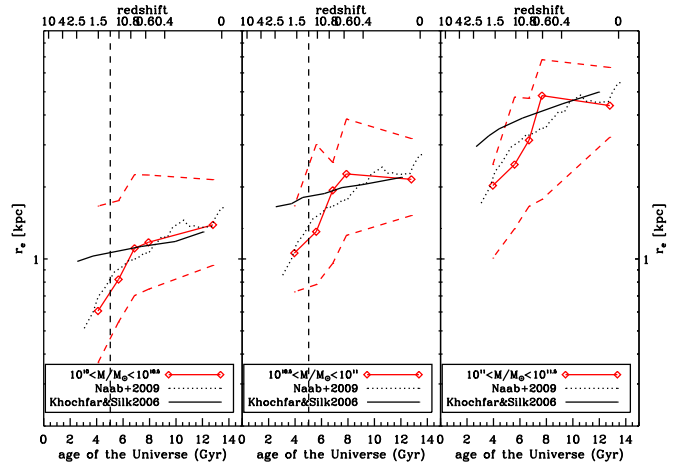}
\caption{Evolution of the average size for passive galaxies at
  $0<z<2.5$, for 3 mass intervals: $10^{10}<M/M_{\odot}<10^{10.5}$
  (left), $10^{10.5}<M/M_{\odot}<10^{11}$ (center) and
  $M/M_{\odot}>10^{11}$ (right). The dashed lines represent the
  $1-\sigma$ scatter, i.e. they contain 68\% of the objects. The
  dotted line is the size evolution for a galaxy with a final mass at
  $z\sim0$ of $M/M_{\odot}\sim10^{11}$ that grows via minor merger
  modeled via a high-resolution hydrodynamical simulation by
  Naab~et~al.~(2009), and the continous line shows a semi-analytical
  model by Khochfar\&Silk~(2006). The datapoints left of the
  vertical line at $z=1.2$ are not complete in the first two mass
  regimes. }\label{fig6}
\end{center}
\end{figure}


In Figure~\ref{fig4} we show the mass-size relation for the 52
galaxies at $1.2<z<2.5$ selected in the ERS+HUDF WFC3 fields, in the
same three redshift bins of Figure~\ref{fig2}.  The sizes have been
measured by GALFIT in the F160w filter, that at these redshifts
corresponds to the optical rest-frame regime. We plot as well the
mass-size relation for local ETGs drawn from the SDSS, in the same
optical rest-frame. We stress that here the mass measurements were
homogenized to a Salpeter IMF (see Section 2 for details). The
relation for SDSS passive galaxies is found to be in good agreement
with results by Shen~et~al.~(2003), that are widely used by many
authors as a reference at $z\sim0$. At all redshifts, we call {\it
ultra-compact} any early-type galaxies 0.4 dex smaller than local SDSS
galaxies of the same mass, and {\it compact} any galaxy $1-\sigma$
below the local relation.  The first definition comes out naturally as
the maximum size without a counterpart in the local universe (see
Section~5); the second is the least strict definition of {\it
compactness} used in the literature and in this way we can compare our
measurements with others. The {\it normal} ETGs are those lying ontop
or above the local relation. If we define the mean mass density within
the half light radius as:
\begin{equation}
\Sigma_{50}=\frac{0.5 M_*}{\pi r_e^2}
\end{equation}
we note that the local relation is roughly parallel to the locii of
constant $\Sigma_{50}$. Compact and ultra-compact ETGs have mass
densities $\Sigma_{50}\gtrsim3\times10^{9}M_{\odot} kpc^{-2}$ and
$\Sigma_{50}\gtrsim1.2\times10^{10}M_{\odot} kpc^{-2}$, respectively.

For each redshift bin we show the region of the plane where our
selection detects at least 80\% of the objects. To build this region
we took the reddest template available in our sample, for which we
know, at every redshift, the $z-$band magnitude as a function of the
stellar mass (see Fig.~\ref{fig2}). We combine this with the output of
the Montecarlo simulation that we used to assess the completeness of
the $z-$band detection images: in particular, we know the fraction of
retrieved objects as a function of the magnitude $m_z$ as well as of
the half-light radius $r_e$. Finally, we built, at each redshift bin,
the locus of the 80\% completeness limit in the stellar mass vs $r_e$
plane. The detection fraction at smaller masses or larger sizes of
this model drops really quickly: in practice, objects left of this
line can not be detected by our selection method (the few objects
lying out of the forbidden region are bluer than the extreme template
we used for this exercise); on the other hand, this work ensures that
we detect at least 80\% of the objects lying to the right of the
line. At $1.2<z<1.6$, our sample is complete down to
$M=10^{10}M_{\odot}$ (as already discussed in section~2,
fig.~\ref{fig2} and up to $r_e\sim3\--10$ (depending on the mass). At
$1.6<z<2$ the limit moves to higher mases, but still we can detect
80\% of the galaxies with $M>10^{10.4}M_{\odot}$ and sizes
$r_e\sim3\--10$. Finally, at $2<z<2.5$ we can detect 80\% of the
galaxies with $M>10^{10.8}M_{\odot}$ and sizes $r_e\sim3\--10$.


The first interesting result is that at $1.2<z<1.6$, $1.6<z<2$
and $2<z<2.5$ respectively about 65\%, 95\% and 100\% of early-type
galaxies are more than $1-\sigma$ below the local relation, and that
about 25\%, 50\% and 80\% are ultracompact, according to our
definition.  The evolution of these fractions between $1.2<z<2.5$
seems to be real: in fact, even considering only the part of the plane
where the detection is higher than 80\%, the result still holds. In
particular, at $M\gtrsim10^{10.6-8}M_{\odot}$, we should detect
``normal'' ETG with $r_e\sim2-4$ kpc at $z>1.6$, if they were present,
but we detect none. However, since at $z>1.6$ we might progressively
miss more and more large galaxies, it is possible that the compact and
ultra-compact fractions at those redshifts are slightly
over-estimated. In the following sections and figures, in order to
improve the statistics, we will combine the three redshift bins of
Figure~\ref{fig4} in one. Note that the absence in our sample of very
massive galaxies with large size is due to their rarity, as with a
surface density of one every $\sim 500$ arcmin$^2$
(Mancini~et~al.~2010) one does not expect to find any in the 40
arcmin$^2$ covered by the WFC3 data.


At $1.2<z<2.5$, passive galaxies appear to be on average 3--5 times
smaller than local counterparts of the same mass. This result is in
good agreement with our previous findings in Cassata~et~al.~(2010) and
with Ryan~et~al.~(2011), obtained with WFC3 in the ERS and/or HUDF
surveys. This result is also in agreement with previous studies at
high$-z$ using NICMOS data (Toft~et~al.~2007; Zirm~et~al.~2007;
Van~Dokkum~et~al.~2008; Buitrago~et~al.~2008) and ACS data
(Trujillo~et~al.~2007; Cimatti~et~al.~2008). However, we do not find
evidence for a population of ``normal'' passive $z\sim2$ galaxies as
abundant as the one presented by Saracco~et~al.~(2009; 2010): they
claim that $\sim$60\% of all early-types at $1<z<2$ are indeed ontop
of the local relation.  This discrepancy may in part be due to the
slightly different redshift interval they use: if we select passive
galaxies in the same redshift interval used by Saracco~et~al.~(2010),
$0.9<z<1.92$, we find a fraction of compact galaxies of
$\sim$60\%. Moreover, we note that Saracco~et~al.~(2009) claim that
the {\it normal} early-types at $z\sim2$ (i.e. those lying on the
local relation) have much younger ages than the {\it compact} ones. It
is possible that we do not select such young early-type galaxies
because of our very conservative SSFR selection ($SSFR<10^{-2}
Gyr^{-1}$). We also note that Mancini et al. (2010) find a
predominance of {\it normal}-size ETGs in a sample of 12 very massive
such galaxies at $1.4<z_{\rm phot}<1.7$, with $M>2.5\times 10^{11}\,
M_\odot$ (Chabrier IMF), a mass range which is not reached by the
present sample.


We thus confirm that passive galaxies at $z\sim2$ are on average
smaller that local counterparts, at least in the mass range
$10^{10}<M/M_{\odot}<10^{11}$ in which our sample is representative,
but we also confirm that compact and normal-size galaxies coexist up
to this high redshift. The aim of this work, however, is to follow the
global evolution of the passive galaxies in the mass-size plane from
$z\sim2.5$ to $z\sim0$, using for the first time an homogeneous and
complete dataset. The evolution at $z<1.2$ is illustrated by
Fig.~\ref{fig5}, where we show the mass-relation for 4 redshift bins:
$0.3<z<0.68$, $0.68<z<0.95$ and $0.95<z<1.2$ and $1.2<z<2.5$ (the
fourth combines together the 3 redshift bins of Figure~\ref{fig4}). It
is evident that the average mass-size relation evolves continuously
from $z\sim2$ to $z\sim0$.  This migration is almost completed by
$z\sim0.4$, as shown in the first panel of Fig.~\ref{fig5}, where the
distribution of galaxies is somehow overlapping with the $z\sim0$ SDSS
relation (although the scatter is larger for the $0.3<z<0.68$
galaxies). The fraction of ``compact'' galaxies, defined as the
galaxies that at each redshift lie $1-\sigma$ below the local relation
(the gray strip in Fig.~\ref{fig5}) drops from $z\sim2$ to $z\sim0$
following such evolution, as we will see in more detail in the next
Section.

Figure~\ref{fig6} shows the evolution of the average size of passive
galaxies between $z\sim2$ and $z\sim0$ for three different mass
intervals: $10^{10}<M/M_{\odot}<10^{10.5}$,
$10^{10.5}<M/M_{\odot}<10^{11}$ and $M/M_{\odot}>10^{11}$. We stress
that at $1.2<z<2.5$ the sample is only complete down to
$M/M_{\odot}=10^{10.8}$.

It is clear, once again, that the evolution is faster at $z>0.6$ (or
age of the universe $<6-8$ Gyr), while at $z<0.6$ the average sizes
remain almost unchanged. Our results are in quantitative agreements
with Stott~et~al.~(2011), who claimed that the average size of the
most massive ETGs increases by 30\% from $z\sim1$ to $z\sim0$. The
shape of the evolution, in a logarithmic plane, is similar for the
three mass bins, so the {\it fractional} size increment as a function
of the cosmic time appears to be independent of the stellar mass. This
implies that on a linear scale the size evolution is faster for the
most massive galaxies than for less massive ones, in agreement with
Ryan~et~al.~(2011).

We note also that the evolution of the average size from $z\sim2$ to
$z\sim0$ is in qualitative agreement with the prediction by
Naab~et~al.~(2009), who modeled the formation and evolution of a
passive galaxy with a final mass of $1.5\times10^{11}M_{\odot}$ in a
high-resolution hydrodynamical simulation. In such simulation, the
mass of the galaxy increases by a factor of 2 from $z\sim2$ to
$z\sim0$ through a series of minor merger events, that at the same
time are responsible for the decrease of the stellar density and
increase of the size throughout the same redshift interval. The
observed size evolution is not well reproduced by the model by
Khochfar\&Silk~(2006), who used a semianalytical technique in which
the size evolution is the result of the amount of cold gas available
during major merger events.

\section{Evolution of the number and mass density of passive galaxies}

\begin{figure}
\begin{center}
\includegraphics[width=1.05\columnwidth]{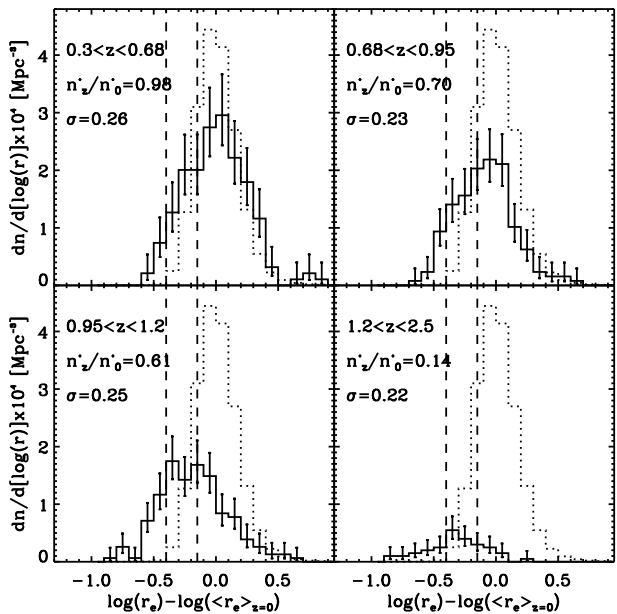}
\caption{For each of the four redshift bins of Figure~\ref{fig6} we
  report the histograms of the ETGs sizes. In particular, for each
  galaxy we calculated the ratio between its size and the average size
  of local ETGs of the same mass. The histograms are normalized to the
  number density of ETGs in each redshift bin (solid line). The dotted
  line is the size histogram for early-type galaxies in SDSS. The
  error bars reflect the Poisson noise in each size bin. The vertical
  dashed lines at $\log(r_e)-\log(<r_e>_{z=0})=-0.15$ and
  $\log(r_e)-\log(<r_e>_{z=0})=-0.4$ indicate our definition of {\it
  compact} and {\it ultra-compact} ETGs given in Section~2. In each
  panel, $n^*_z$ and $n^*_0$ indicate the number density of ETGs in
  that redshift bin and in the local Universe, respectively.  For each
  panel, $\sigma$ indicates the standard deviation of the size
  distribution, to be compared with $\sigma=0.17$ for the SDSS $z=0$
  galaxies. }\label{fig7}
\end{center}
\end{figure}

\begin{figure}
\begin{center}
\includegraphics[width=\columnwidth]{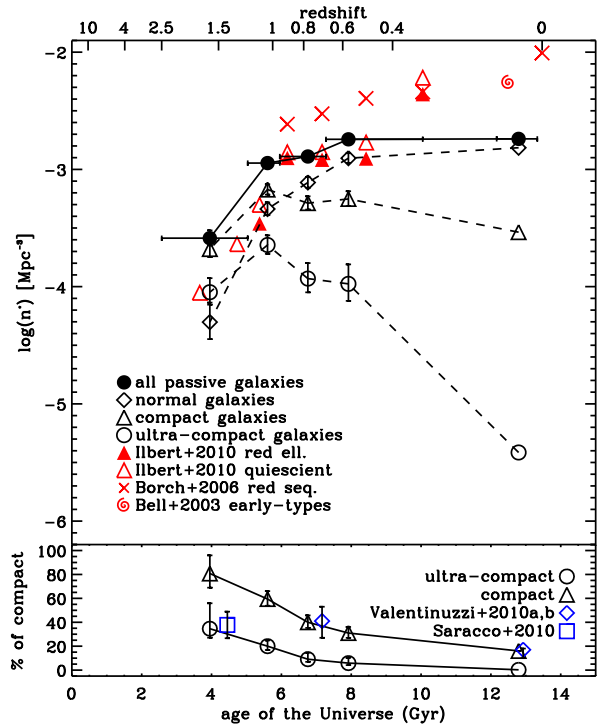}
\caption{{\it Top panel:} number density as a function of the age of
  the universe and redshift of passive galaxies with
  $M>10^{10}M_{\odot}$. Filled circles indicate the mass density for
  all massive passive galaxies in our sample; the empty triangles
  represent the mass density in compact passive galaxies, defined as
  the objects lying $1-\sigma$ below the local SDSS relation (or
  equivalently with $\Sigma_{50}>3\times10^9M_{\odot}$); empty circles
  represent ultra-compact galaxies, those galaxies 0.4 dex smaller
  than local counterparts; empty diamonds indicate the number density
  in normal passive galaxies, i.e. galaxies that reside on the local
  mass-size relation. The points at $z\sim0.1$ have been derived using
  the local SDSS sample defined in Section~2. Red symbols are
  literature results at different redshifts, obtained integrating the
  best fit Schechter functions down to $M=10^{10}M_{\odot}$: filled
  and open triangles are results by Ilbert~et~al.~(2010) for red
  ellipticals and quiescent galaxies (SSFR$<10^{-2}Gyr^{-1}$),
  respectively; red crosses are measurements by Borch~et~al.~(2006)
  for galaxies in the red sequence; the red spiral indicates the local
  value estimated by Bell~et~al.~(2003) in the local Universe for
  early-type galaxies, defined as the objects with concentration
  $c_r>2.6$.  {\it Bottom panel:} Fraction of passive galaxies that
  are also compact (triangles) and ultracompact (circles), as a
  function of the age of the universe or the redshift.
}\label{fig8}
\end{center}
\end{figure}

\begin{figure}
\begin{center}
\includegraphics[width=\columnwidth]{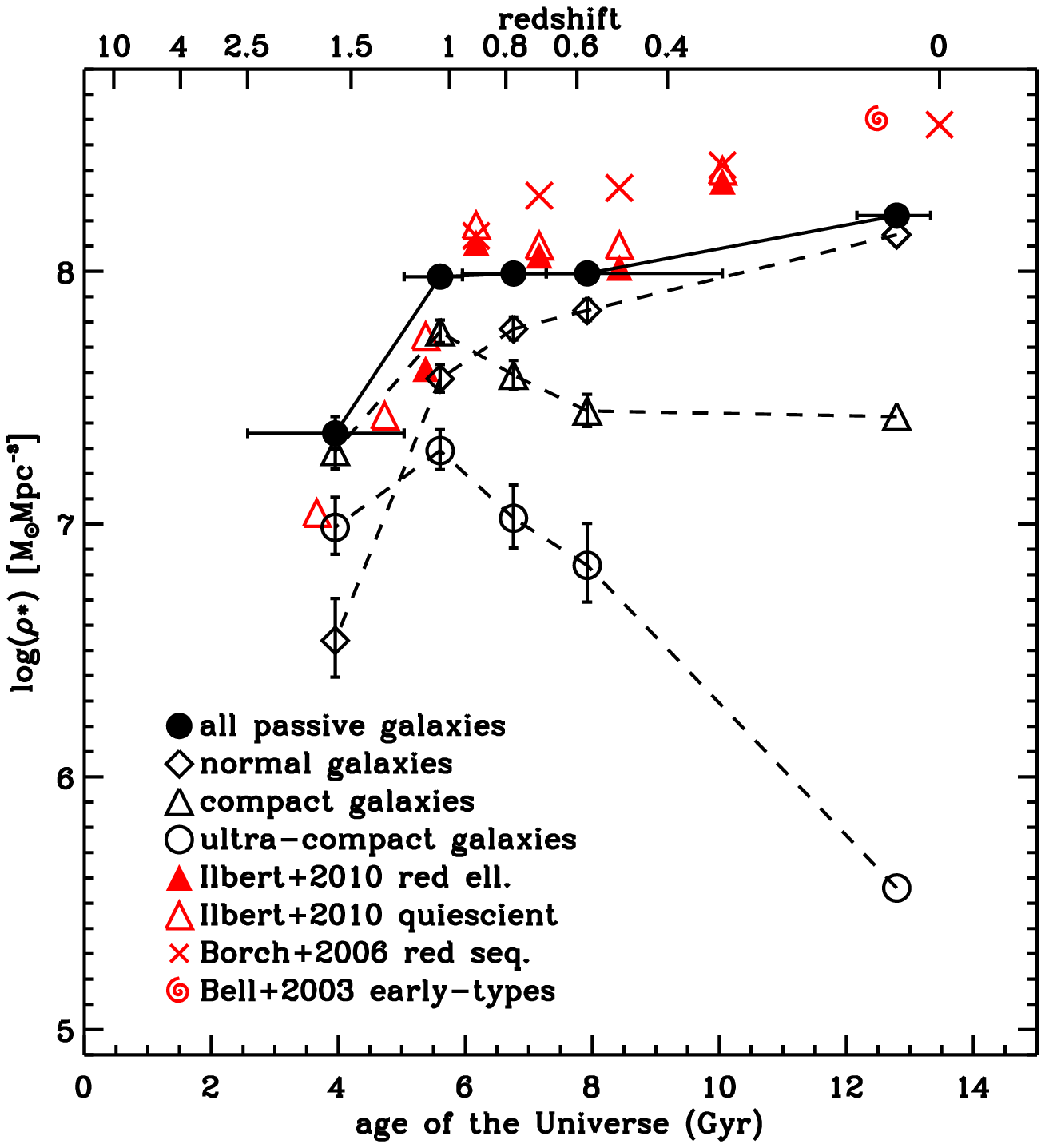}
\caption{Stellar mass density as a function of the age of the universe
  and redshift of passive galaxies with $M>10^{10}M_{\odot}$. The
  symbols are the same as in Figure~\ref{fig8}}\label{fig9}
\end{center}
\end{figure}

In the previous section we have observed that the average mass-size
relation of passively evolving early-type galaxies undergoes
significant evolution from $z\sim2$ to the present.

In principle, one can imagine that the evolution is driven by two
independent mechanisms: one is the progressive appearance, as time
evolves, of new early-type galaxies due to the cessation of the
star-formation activity; the other is the evolution of galaxies that
are already passive, and whose stellar populations remains mostly
passive, but whose mass and size change ad the result of merging and
interactions with other galaxies, passive or otherwise. Some models
including these effects have been proposed recently:
Khochfar\&Silk~(2006) suggested that the observed size evolution of
passive galaxies can be explained by the variation of the amount of
cold gas available during the major merger events that originate
early-type galaxies: the most massive ones formed at high-$z$, when
major mergers were more gas rich than at later
epochs. Naab~et~al.~(2009), on the other hand, introduced a model in
which the size of a passive galaxy increases, as time goes by, through
the continous merging activity with low mass objects.

At any redshift, the appearance of new passive galaxies modifies the
mass-size relationship depending on the mass and size distribution of
the galaxies that become passive. On the other hand, the morphological
transformation of the old passive galaxies contributes to modify the
mass--size relationship depending on how galaxies accrete stellar mass
and/or grow in size, for example either increasing stellar mass but
remaining unchanged in size, or also growing by ``puffing-up''
(e.g. Hopkins~et~al.~2009).

Van~der~Wel~et~al.~(2009) found that in the local universe small
passive galaxies with $M>10^{11}M_{\odot}$ are older than the larger
ones, implying that at higher redshift such larger galaxies gradually
disappear. They suggest that this effect, together with some dry
merging activity, may account for the observed size evolution from
$z\sim2$ to $z\sim0$.

In this paper, we document the evolution of the mass-size relation in
terms of the relative importance of the size growth of individual
passive galaxies and the appearance of new large ETGs at more recent
epochs. To this purpose, we constrain the number density of ETGs for
different sizes as a function of the redshift. We stress that up to
date the evidence that compact early-type galaxies are more common at
high-$z$ than in the local Universe is mainly based only on the fact
that their ``fraction'' increases with $z$, irrespectively of the fact
that the total number of ETGs is not constant throughout the life of
the Universe (e.g. Trujillo~et~al.~2009). Nevertheless, some authors
(Valentinuzzi~et~al.~2010a,b; Taylor~et~al.~2010; Saracco~et~al.~2010)
have compared the {\it number density} of compact ETGs at $z\sim2$ and
$z\sim0$, but they used non homogeneous samples of ETGs selected with
different techniques at high- and low-$z$. In this work, for the first
time, we select in an homogeneous way a mass complete sample of ETGs
and we measure the number density of compact and normal ETG throughout
the entire $0<z<2$ redshift range.

We used a simple $V/V_{max}$ formalism to measure the number and mass
densities in early-types with $M>10^{10}M_{\odot}$ in the same four
redshift of Figure~\ref{fig5} and for the local SDSS sample at
$z\sim0$. Our sample is complete down to this mass limit just to
$z=1.6$, while at $z>1.6$ the sample is only complete down to
$M=10^{10.8}M_{\odot}$ (see Fig.~\ref{fig2}). The $V/V_{max}$,
however, takes into account such incompleteness and corrects for it.
In particular, our method assumes that the density of galaxies in
the yellow region of Fig.~\ref{fig2} is the same as the one that we
measure for galaxies with $m_z<26.5$.


In Figure~\ref{fig7} we show the size distribution of ETGs in the same
four redshift bins as Figure~\ref{fig5}, scaled to the number density
of galaxies in each bin and compared with the distribution of galaxies
in the local universe derived from our SDSS sample. For each galaxy in
the GOODS and SDSS sample we calculated the ratio of its size to the
average size of SDSS ETGs with the same stellar mass, and then we
plotted the histograms of such ratio. The local size distribution is
almost perfectly lognormal, with virtually no {\it ultra-compact} ETGs
(e.g. galaxies with sizes 0.4 dex smaller than the local
average). However, some authors suggested that a compact population of
ETGs could be hidden among unresolved red objects in SDSS, and thus
their number density could be somewhat underestimated
(Valentinuzzi~et~al.~2010a; Shih~\&~Stockton~2011). The size
distribution at higher redshift is approaching a lognormal as well,
eventhough it is broader than the $z\sim0$ one. The standard deviation
of the distribution is in fact $\sigma\sim0.22\--0.26$ at $0.3<z<2.5$,
while it is $\sigma\sim0.17$ at $z\sim0.$ This effect might be due to
the fact that $z>0.3$ size measurements are noisier than the $z\sim0$
ones, thus producing a broader distribution.



The bottom right panel of Fig.~\ref{fig7} shows that at the same
stellar mass ETGs at $1.2<z<2.5$ are on average 0.5 dex smaller than
their local counterparts. This indicates that early-type galaxies
assembled at those epochs are preferentially small. Note that we are
not suggesting that the size distribution at high redshift is
truncated: galaxies with large size become rarer at $z>1.2$, implying
that surveys with larger area coverage than ours are required to find
them in significant numbers (e.g. Mancini et~al. 2010).

By redshift $z\sim1$, the number density of ETG has increased by a
factor of 5 (see also Figure~\ref{fig8}). We note that the number
density of ETGs increased significantly at all sizes from $1.2<z<2.5$
to $0.95<z<1.2$: if for a moment assume that the galaxies already
present at $z>1.3$ do not grow in size, the ETGs formed during this
epoch would have a size distribution spanning from -0.5 dex smaller to
0.5 dex larger than local counterparts. This indicate that the new
galaxies formed over this period have typically larger sizes
than those already present at $z>1.2$. Even though the size
distribution in the $0.95<z<1.2$ bin is significantly different than
the local one, the number density of galaxies 0.3 dex larger than
local counterparts is now similar to the local value.

The evolution at later epochs is much milder, with the histogram
progressively skewing towards larger sizes, and with the total number
density of ETGs increasing by a factor of $\sim$1.5. The ultra-compact
objects, i.e. those with size 0.4 dex smaller than local counterparts,
gradually disappear, while the number of galaxies with sizes
comparable to the local value ($\log(r_e)-\log(<r_e>_{z=0})\sim0$)
steadily increases. This indicates that in this phase the smallest
galaxies grow in size, and that the new galaxies formed have sizes
comparable to the local counterparts. We note that the evolution in
the lowest redshift bin is not yet complete: even though the peak of
the histogram is close to the peak of the local SDSS one, and the tail
at large size is quite similar to the local one, still some
ultra-compact galaxies are present. In the next 5 Gyr, i.e. between
$z\sim 0.5$ and $z\sim 0$, they keep growing in size and reach the
present distribution, with essentially negligible new ETGs added (the
area enclosed by the two histogram is almost the same, see also
Figure~\ref{fig8}). As already noted before, the size distribution at
$z\sim0$ is much narrower than that at $0.3<z<2.5$: this can be in
part due to the errors on the size measurements at high-$z$. However,
while the distribution at large sizes is quite similar at $z>0.3$ and
$z\sim0$ (and the small difference can be explained with a 5\% of
outlyers), the excess of small galaxies at $z>0.3$ is much more
significant, and, if not true, would imply a unlikely large number of
small size outlyers (50\% of the objects at $z>1.2$ and 15--20\% at
$0.3<z<1.2$.

Figures~\ref{fig8} and \ref{fig9} show the evolution of the number
density $n^*$ and stellar mass density $\rho^*$ contributed by normal,
compact and ultra-compact ETGs with $M>10^{10}M_{\odot}$ as a function
of redshift and cosmic time, together with the evolution of the
fraction of passive galaxies that are also compact and ultra-compact
(bottom panel in Figure~\ref{fig8}). The data at \zmed$=0.1$ have been
derived using the SDSS local sample defined in Section 2. Our high-
and low-$z$ samples probe $\sim6$ Gyr of cosmic time, from the epoch
when the Universe was 4 Gyr old to the epoch it was 10 Gyr
old. Including the point at $z\sim0$, we can follow the evolution up
to 13.6 Gyr after the Big Bang.


In order to estimate a reliable error to assign to the mass and number
density at each redshift we used a simple Monte Carlo simulation. In
particular, we built 1000 realizations of our sample by perturbing the
stellar mass of each galaxy assuming a typical (quite conservative)
random error of $\pm$0.4 dex, and we thus measured 1000 realizations
of the mass density as a function of the redshift.  In
Figs.~\ref{fig8} and \ref{fig9} we use the median and standard
deviation of such 1000 measurements as the value of the mass density
and relative error, that thus do not include any estimate of the
cosmic variance.



We compared our measurements with results in the literature for
samples of similarly selected passive galaxies. In particular, we took
the best fit Schecther parameters by Ilbert~et~al.~(2010),
Borch~et~al.~(2006) and Bell~et~al.~(2003), we homogenized their
masses to our assumptions and we integrated their Schechter function
down to our mass limit ($M_*>10^{10}M_{\odot}$), in order to have
homogeneous measurements. However, since the mass functions of
passively evolving galaxies have all moderate slopes at low mass
values, both the number density $n^*$ and the mass density $\rho^*$
are dominated by galaxies around $M^*$, that is typically $M^*\simeq
10^{10.6}M_{\odot}$ (Ilbert et al. 2010; Peng et al. 2010). Thus, the
contribution to the the number and mass density of galaxies with
$M^*<10^{10}M_{\odot}$ is negligible, and the values presented here
can be considered total values. The measurements by
Saracco~et~al.~(2010) are instead for galaxies with
$\sim4.5\times10^9M_{\odot}$, but again this different limit should
not affect our comparison.  The incompleteness of our sample at
$z>1.2$ do not strongly affect the conclusions of this
discussion. Even if we eliminated all galaxies with $z>1.6$ (the
redshift at which the sample is complete down to $M=10^{10}M_{\odot}$),
we would have found a comparable number density of ETGs.

Our results for the global population of passive galaxies at
$0.5<z<2.5$, both in Figure~\ref{fig8} and \ref{fig9} are in fair
agreement with the results for quiescent and red elliptical galaxies
by Ilbert~et~al.~(2010). We do see, however, a gap between our
measurements at $z<1$ and those by Borch~et~al.~(2006) and
Bell~et~al.~(2003). These differences are probably due to the
different selection criteria: our selection is very conservative, and
just at all redshifts selects galaxies with a tight upper limit to
their SFR, and an early-type morphology, whereas Borch~et~al.~(2006)
refer to the mass density for all galaxies on the red sequence, that
can be easily contaminated by dusty star forming galaxies
(Cassata~et~al.~2008). On the other hand, Bell~et~al.~(2003) select as
early-types the galaxies with concentration in the $r-$band $c_r>2.6$,
that can thus be contaminated by bulge-dominated spirals.



By comparing Figure~\ref{fig8} and Figure~\ref{fig9} we can draw
interesting conclusions on the evolution of the passive population
from \zmed$=1.6$ to $z\sim0$. In particular, both the mass density and
the number density of early-type galaxies increases by a factor of
$\sim$5 from \zmed $=1.6$ to $z\sim1$, similarly to what has been
found by Franceschini~et~al.~(2006), Arnouts~et~al.~(2007) and
Abraham~et~al.~(2007). This indicates that the bulk of the quenching
of star formation if such galaxies takes place around that epoch. The
evolution at later epochs is much milder. While the number density
increases by a factor of 1.5 between $z\sim1$ and $z\sim0.5$, the mass
density remain constant over the same interval. This implies that new
early-type galaxies are constantly added from $z\sim1$ to $z\sim0.5$,
as already seen in Figure~\ref{fig7}, but they are not massive enough
to modify significantly the mass density. In other words, the new
early-type galaxies added from $z\sim1$ to $z\sim0.5$ are low mass
objects, in good agreement with the ``downsizing'' scenario for the
evolution of such galaxies (e.g., Thomas et al. 2010).  We note that
the data by Ilbert~et~al.~(2010) show a similar behavior.

However, we can not draw any strong conclusions, based on our
data, on the evolution of the number density and mass density of ETGs
from $z\sim0.5$ (the smallest GOODS redshift bin) and $z\sim0$ (the
local SDSS value), since the GOODS fields are too small to minimize
the effects of the cosmic variance.


Figures~\ref{fig8} and \ref{fig9} distinguish the number and density
evolution of compact, ultra-compact and normal ETGs as a function of
redshift. In particular, Figure~\ref{fig8} shows that the number
densities of the {\it compact} and {\it ultra-compact} ETGs are not
constant through the explored redshift range. We stress that at
$z\sim0$ the classical definition of compactness sets to 16\% the
fraction of {\it compact} galaxies. At first, the number densities of
compact and ultra-compact ETGs increase following the the strong
evolution the global passive population: by $z\sim1$ there appears to
be $\sim$3~times more compact and ultra-compact galaxies than those we
found at \zmed$=1.6$.  At $z<1$, conversely, their number density
slowly decreases with cosmic time, decoupling from the evolution
observed for the general population. At $z\sim0$, the number density
of {\it compact} ETGs has decreased to 16\% of the total density (by
definition), reaching a value that is, just by chance, comparable to
the number density at \zmed$=1.6$
(e.g. Saracco~et~al.~2010). Similarly, the number density of the {\it
ultra-compact} ETGs drops by a factor of $\sim$100 from z$\sim1$ to
$z\sim0$, and in the local universe ultra-compact ETGs are $\sim$500
times rarer than {\it normal} ETGs. Similar behaviors are exhibited by
the mass density in compact ETGs as a function of the redshift in
Figure~\ref{fig9}.

It is interesting to note that in Figure~\ref{fig8} the number of {\it
  normal} early-types from $z\sim1$ to $z\sim0.5$ increases at a
faster rate than the one at which the {\it compact} ones disappear. As
a result, the total number density increases, as already
described. This result implies once again that both the mechanisms
mentioned at the beginning of this section are at work from $z\sim1$
to $z\sim0$. On one side, some of the {\it compact} galaxies gradually
increase in size, moving onto the local relation, and causing number
(and mass) density of compact early-types to decrease; the {\it
  compact} galaxies that become {\it normal} support in part the
growth of the mass and number density of {\it normal} early-types; but
since the total number (compact+normal) increases as well, it means
that new passive galaxies with sizes comparable to local counterparts
are formed.

In the bottom panel of Fig.~\ref{fig8} we show the fraction of compact
and ultra-compact ETGs as a function of the redshift and cosmic
time. The fraction of compact ETGs decreases from 80\% at \zmed$=1.6$
to 16\% at $z=0$, while the fraction of ultra-compact drops from 40\%
at \zmed$=1.6$ to zero at $z=0$. We stress that these fraction
can be slightly overestimated at $1.2<z<2.5$, as a result of the
incompleteness of our selection at $z>1.6$. Anyway, strong lower
limits to these fractions (if we limit the sample to $1.2<z<1.6$) are
set to 65 and 25\%, respectively. We note again that
Saracco~et~al.~(2010) found a much smaller fraction of compact
early-types at $z\sim2$, and we tried to explain such difference in
the previous Section. However, our estimates for the compact fraction
are in quite good agreement at later epochs with the measurements by
Valentinuzzi~et~al.~(2010a,b). In particular,
Valentinuzzi~et~al.~(2010a) analyzed a sample of $z\sim0.04$ massive
galaxies in clusters (WINGS survey, Fasano~et~al.~2006), and found a
significant fraction of very compact
galaxies. Valentinuzzi~et~al.~(2010b) found a similar rich population
of compact galaxies in clusters at $z\sim0.7$ (EDisCS,
White~et~al.~2005), and claimed that 17\% and 44\% of all cluster
galaxies are compact in WINGS and EDisCS clusters, respectively. By
knowing the fraction of galaxies in the two surveys that are not
morphologically early-type, we could reconstruct the fraction of
compact early-types that we report in the bottom panel of
Fig.~\ref{fig8}.


\section{Discussion and conclusions}
In this paper we have analyzed a complete and homogeneous sample of
passive galaxies from $z\sim0$ to $z\sim2.5$, with the aim of
observationally constrain the mass and size evolution of such systems
during the last 10 Gyr of cosmic time. In particular, we estimated the
evolution of the relative abundance of {\it compact}, {\it
  ultra-compact} and {\it normal} ETGs throughout this redshift range.

We selected a homogeneous and robust sample of 563 passive galaxies
using the multiband photometry available in the GOODS N+S fields,
selecting galaxies with $M>10^{10}M_{\odot}$ and
$SSFR<10^{-2}Gyr^{-1}$. We studied the morphological properties of the
sample in the optical rest-frame, using the ACS $z-$band for objects
at $z<1.2$ and the WFC3 $H-$ band for galaxies at $z>1.2$, and we
discarded galaxies with late-type morphologies. We then run GALFIT to
measure the S\'ersic indices and physical sizes.

We found that, at all redshifts, the morphology traced by the
rest-frame UV light is, within the statistical uncertainties, very
close to that traced by the optical light, at least when parameters
such as the S\'ersic indices and effective radii are used. In fact,
the sizes and S\'ersic indices measured in the UV and optical
rest-frame correlate quite well with each other, with the sizes
measured in the optical rest-frame being slightly smaller (by
$\sim$20\% and $\sim$10\% at $z\gtrsim1$ and $z\lesssim1$,
respectively) than those derived from the UV rest-frame. This shows
that, at least for galaxies that are not currently forming stars, the
morphological K-correction is weak, and thus that ACS/$z-$band
observations, if deep enough, can be indeed used to measure the size
of passive galaxies at $z>1$. This is in good agreement with previous
results by Trujillo~et~al.~(2007) and Cassata~et~al.~(2010) and, at
lower redshift, with Cassata~et~al.~(2005) and Papovich~et~al.~(2003).

We showed that the mass-size relation of ETGs evolves significantly
from \zmed 1.6 to $z\sim0$: $\sim 80\%$ of ETGs at \zmed$=1.6$ in our
sample are smaller than local counterparts of the same mass, and at
later epochs ETGs are increasingly larger. We showed as well that the
fractional increase of the average size is almost independent on the
stellar mass.

With the aim of understanding whether the evolution of the mass-size
relation is driven by the size growth of each individual galaxy or by
the appearance of new large ETGs, we have built the size distribution
in four redshift bins, scaled to the number density of ETGs in each
bin. We observed that ETGs at $1.2<z<2.5$ are on average 0.5 dex
smaller than local counterparts of the same mass, and we identified a
rich population of {\it ultra-compact} ETGs that have no identified
counterparts in terms of size and density in the local Universe. As
the time goes by, at $0.95<z<1.2$ the number of ETGs is increased by a
factor of 5, and the new ETGs have a broad range of sizes, from 0.5
dex smaller to 0.5 dex larger than local counterparts of the same
mass. At $z<0.95$ the evolution is milder, with the number of ETGs
increasing of another factor of 1.5 down to $z\sim0$; the sizes of the
new ETGs formed in this phase are similar to the average local value;
at the same time the small ETGs gradually disappear, growing in size
and skewing the size distribution towards large sizes. These results
show that both the mechanisms driving the evolution of the mass-size
relation are at work at the same time.


We measured the mass and number density of early-type galaxies as a
function of redshift, including the evolution of the relative fraction
of {\it normal}, {\it compact} and {\it ultracompact} ones. More in
details, we showed that the number and mass density of massive and
passive early-types increase by a factor of $\sim 5$ within the $\sim
2$ Gyr between \zmed$=1.6$ to $z\sim1$, in agreement with
Arnouts~et~al.~(2007) and Ilbert~et~al.~(2010). In the following 4
Gyr, from $z\sim1$ to $z\sim0.5$, the number density increases by at
most another factor of $\sim 1.5$, while in the same interval the mass
density remains constant. This implies that, as expected in a
``downsizing'' scenario, the added galaxies that drive the number
evolution have small masses and do not contribute too much to the mass
density. These findings are in good agreement with
Franceschini~et~al.~(2006), Borch~et~al.~(2006) and
Abraham~et~al.~(2007), and similar to the results by
Cimatti,~Daddi~\&~Renzini~(2006) and Scarlata~et~al.~(2007).

From $z\sim1$ to $z\sim0$, we showed also that the number (and mass)
density of {\it compact} early-types decreases by a factor of 2, while
the number density of normal-size ETGs increases much faster,
indicating once again that the overall increase of the average size at
fixed mass is only partly due to a size increase of individual
galaxies, whereas part of the effect is due to the appearance of new
ETGs with large size.


We interpret these results as the evidence that the phases at $z>1$
and $z<1$ are dominated by distinct and different physical
mechanisms. The scenario that comes out is the following: the epoch at
$1<z<3$ is when the bulk of the stellar mass in local early-type
galaxies is assembled. This process of assembly is quite rapid: the
stellar mass locked in passive early-type galaxies increases by a
factor of 5 over a period of $\sim$2 Gyr. Whatever the mechanism
responsible for this huge mass density increase (gas-rich major
merger, collapse of unstable disks, monolithic collapse), it produces
a remnant that is compact and small with respect to local
counterparts: $\sim$80\% of early-types in this redshift interval are
smaller than galaxies of the same mass in the local Universe.
Interestingly, Targett~et~al.~(2011) found that submillimeter galaxies
at $z\sim2$ have similar sizes to those that we measure for ETGs at
the same epoch, suggesting a possible evolutionary connection between
the two classes of galaxies. On the other hand the evolution at $z<1$
is much milder. In this phase, the compact galaxies continuously
increase in size, eventually disappearing, most likely undergoing a
series of dry and wet minor mergers or by slowing accreting a small
amount of mass in their outskirts (e.g. Van~Dokkum~et~al.~2010;
Hopkins~et~al.~2009).  At the same time, new ETGs are formed, driving
the number density increase of a factor 1.5 that we observe from
$z\sim1$ to $z\sim0$. These new ETGs have small masses and large
sizes, indicating that the mechanism through which they are formed is
different from the one that at $z>1$ generated the compact ETGs.



\begin{acknowledgements} 
PC, MG, YG and SS acknowledge support from NASA grants
HST-GO-9425.36-A, HST-GO-9822.45-A, and HST-GO-10189.15-A, awarded by
the Space Telescope Science Institute, which is operated by the
Association of Universities for Research in Astronomy, Inc. (AURA)
under NASA contract NAS 5-26555. The work presented here is partly
based on observations obtained with WIRCam, a joint project of
Canada-France-Hawaii Telescope (CFHT), Taiwan, Korea, Canada, France,
at the CFHT which is operated by the National Research Council (NRC)
of Canada, the Institute National des Sciences de l'Univers of the
Centre National de la Recherche Scientifique of France, and the
University of Hawaii. We are grateful to the anonymous referee
for the useful comments.
\end{acknowledgements}

\end{document}